\documentclass[sigconf]{acmart}
\usepackage{float} 

\usepackage{multirow}
\usepackage{subfigure} 
\usepackage{float}

\usepackage[lined, ruled]{algorithm2e}

\usepackage{hyperref}

\usepackage{listings}

\definecolor{commentcolor}{RGB}{85,139,78}
\definecolor{stringcolor}{RGB}{206,145,108}
\definecolor{keywordcolor}{RGB}{0,0,0}
\definecolor{backcolor}{RGB}{220,220,220}
\definecolor{NavyBlue}{RGB}{0,0,128}
\SetAlCapNameFnt{\small}
\SetAlCapFnt{\small}

\newcommand{\specialcell}[2][c]{\begin{tabular}[#1]{@{}l@{}}#2\end{tabular}}

\newcommand{\eg}{\emph{e.g.}~}
\newcommand{\etal}{\emph{et al.}~}
\newcommand{\ie}{\emph{i.e.}~}
\newcommand{\wrt}{\emph{w.r.t.}~}

\newcommand{\argmax}{\mathop{\rm arg~max}\limits}

\AtBeginDocument{%
  \providecommand\BibTeX{{%
    \normalfont B\kern-0.5em{\scshape i\kern-0.25em b}\kern-0.8em\TeX}}}

\copyrightyear{2022}
\acmYear{2022}
\setcopyright{acmlicensed}\acmConference[MM '22]{Proceedings of the 30th ACM International Conference on Multimedia}{October 10--14, 2022}{Lisboa, Portugal}
\acmBooktitle{Proceedings of the 30th ACM International Conference on Multimedia (MM '22), October 10--14, 2022, Lisboa, Portugal}
\acmPrice{15.00}
\acmDOI{10.1145/3503161.3547968}
\acmISBN{978-1-4503-9203-7/22/10}

\begin{document}
\title{Learn to Understand Negation in Video Retrieval}

\author{Ziyue Wang}
\authornote{Z. Wang and A. Chen contributed equally to this research.}
\affiliation{%
\institution{AIMC Lab, School of Information, Renmin University of China}
\country{China}
}
\author{Aozhu Chen}
\authornotemark[1]
\affiliation{%
\institution{AIMC Lab, School of Information, Renmin University of China}
\country{China}
}
\author{Fan Hu}
\affiliation{%
\institution{AIMC Lab, School of Information, Renmin University of China}
\country{China}
}
\author{Xirong Li}
\authornote{Corresponding author: Xirong Li (xirong@ruc.edu.cn).}

\affiliation{%
\institution{MoE Key Lab of DEKE, Renmin University of China}
\country{China}
}
\renewcommand{\shortauthors}{Wang et al.}

\begin{abstract}
Negation is a common linguistic skill that allows human to express what we do NOT want. Naturally, one might expect video retrieval to support natural-language queries with negation, e.g., finding shots of kids sitting on the floor and not playing with a dog. However, the state-of-the-art deep learning based video retrieval models lack such ability, as they are typically trained on video description datasets such as MSR-VTT and VATEX that lack negated descriptions. Their retrieved results basically ignore the negator in the sample query, incorrectly returning videos showing kids playing with dog. This paper presents the first study on learning to understand negation in video retrieval and make contributions as follows. By re-purposing two existing datasets (MSR-VTT and VATEX), we propose a new evaluation protocol for video retrieval with negation. We propose a learning based method for training a negation-aware video retrieval model. The key idea is to first construct a soft negative caption for a specific training video by partially negating its original caption, and then compute a bidirectionally constrained loss on the triplet. This auxiliary loss is weightedly added to a standard retrieval loss. Experiments on the re-purposed benchmarks show that re-training the CLIP (Contrastive Language-Image Pre-Training)  model by the proposed method clearly improves its ability to handle queries with negation. In addition, the model performance on the original benchmarks is also improved. 
\end{abstract}
\ccsdesc[500]{Information systems~Video search}
\ccsdesc[500]{Information systems~Test collections}


\keywords{Text-to-video retrieval (T2VR), nT2VR, negation learning}


\begin{teaserfigure}
\includegraphics[width=\textwidth]{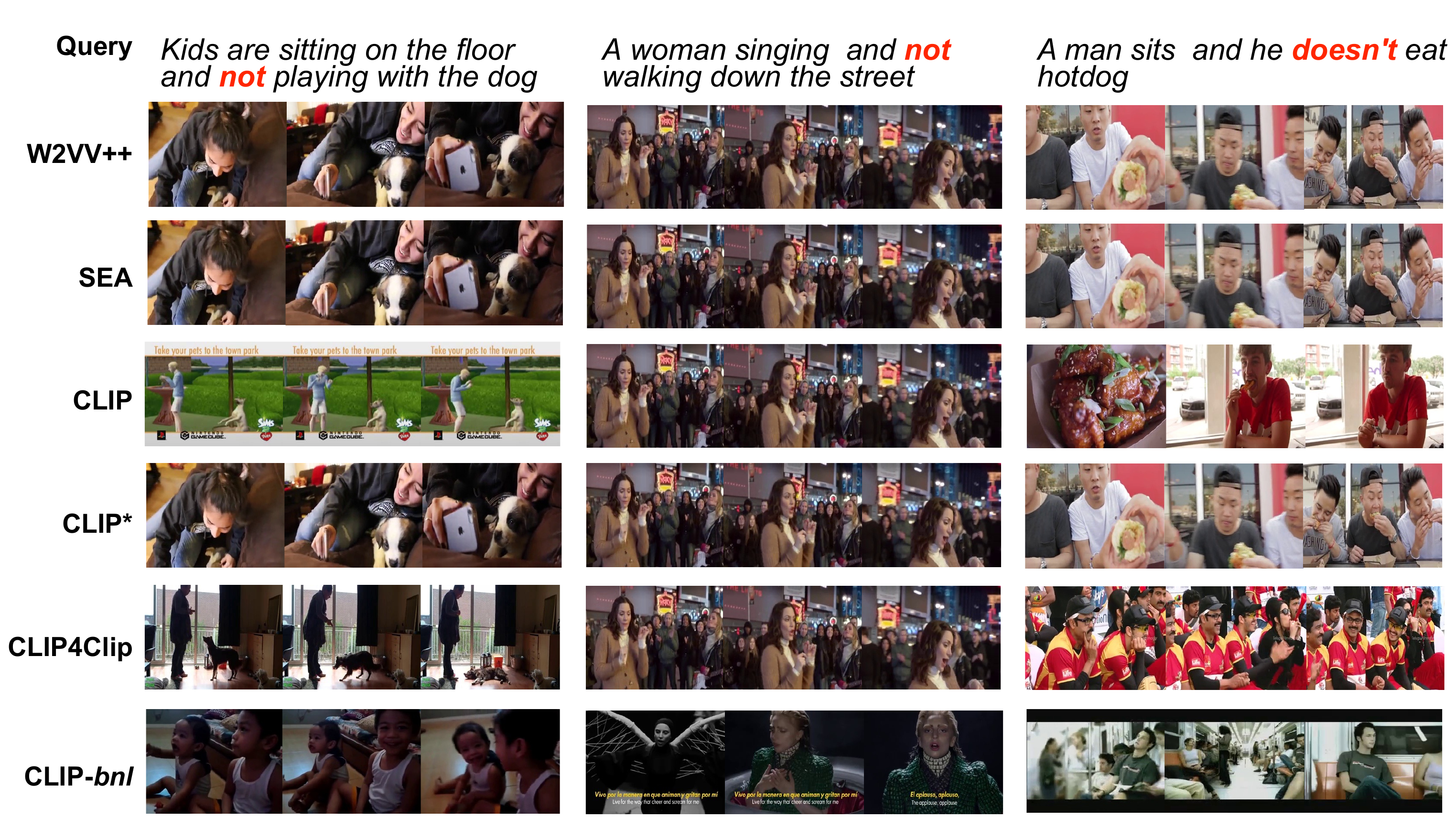}
\caption{\textbf{Top-1 video retrieved by different models}, \ie W2VV++ \cite{LiXirong2019W2VVPP}, SEA \cite{LiXirong2020SEA}, CLIP \cite{2021clip_icml}, CLIP* (fine-tuned by this work), CLIP4Clip \cite{luo2021clip4clip} 
and our CLIP-\emph{bnl}, which is CLIP re-trained with proposed negation learning. This paper presents the first study on a learning based method for handling negation in text-to-video retrieval (nT2VR). Data source: MSR-VTT \cite{MSR-VTT}. } 
  \label{fig:teaser}
\end{teaserfigure}

\maketitle

\section{Introduction}

This paper is targeted at text-to-video retrieval (T2VR), also known as video retrieval by text. T2VR aims to let common users retrieve the increasing amounts of unlabeled videos by textual queries. Due to its high practical value, the topic has attracted much attention recently \cite{icmr18-t2v,LiXirong2019W2VVPP,mm20-video-retrieval,Wu2020InterpretableEF,mm21-hanet,mm21-finegrained}. These dedicated research efforts have been well paid off, with continuous performance improvement reported on both public datasets \cite{MSR-VTT,VATEX} and international benchmark evaluations \cite{ChenHu2021,2021trecvidawad}. 
It seems unquestionable that more powerful T2VR models will be developed to help users  find what they want. The question is do the (current) models understand what the users do \emph{not} want?

\emph{Negation} is an important and common linguistic skill for human beings to express what we do not want. A query with negation can be ``finding shots of kids sitting on the floor and \emph{not} playing with the dog''. As exemplified in Fig. \ref{fig:teaser}, a number of current models, \eg  W2VV++ \cite{LiXirong2019W2VVPP}, CLIP \cite{2021clip_icml} and CLIP4Clip \cite{luo2021clip4clip},
are actually not good at answering this specific query. Recall that these models are trained on video description datasets, such as MSVD \cite{msvd}, MSR-VTT \cite{MSR-VTT} and VATEX \cite{VATEX}, which were originally developed for the video captioning task. For that task, annotators tend to describe what was present in the video content other than what was absent. By a rule-based negation cue detection, we find that only 1.5\% of the MSR-VTT video descriptions contain negation cues such as \emph{no}, \emph{not}, and \emph{without}. . The lack of negated descriptions in the training data has a clear consequence on the models. Their retrieved results basically ignore the negator in the sample query, incorrectly returning videos showing kids playing with the dog, see Fig. \ref{fig:teaser}.


Towards addressing negation in T2VR (nT2VR), an initial attempt has been made by Wu and Ngo \cite{Wu2020InterpretableEF}. The authors describe a rule-based strategy to process queries with negation. In particular, given a query such as \emph{beach not man}, the strategy treats the query as a logic expression of \emph{beach} AND (NOT \emph{man}). The boolean operation is practically implemented by subtracting a video’s relevance score to \emph{man} from its relevance score to \emph{beach}. We consider their study preliminary as they experimented with only five hand-crafted queries. More importantly, the boolean operation is essentially a post-processing trick. The underlying T2VR model remains unaware of the negation.



In this paper, we present a first study on learning to understand negation in T2VR. Our major contributions are as follows: 
\begin{itemize}
    \item Due to the absence of related data and evaluation criteria, we introduce a new evaluation protocol. In particular, we re-purpose MSR-VTT and VATEX by automatically constructing thousands of negated and composed queries from the original descriptions. Such a re-purposing allows the protocol to support large-scale evaluation without the need of extra manual labeling. By preserving the original test queries, the new protocol can also be used to test how a negation-aware model performs on the original benchmarks.
    \item We propose a learning based method for training a negation-aware T2VR model. Specifically, given a training video, its original description and a partially negated description, we compute on the triplet a bidirectionally constrained loss. Consequently, negation learning (NL) is realized with ease by adding this auxiliary loss to a standard retrieval loss.
    \item Extensive experiments on the two re-purposed benchmarks show that re-training the CLIP (Contrastive Language-Image Pre-Training)  model \cite{2021clip_icml} by the proposed method clearly improves the model's ability to handle queries with negation. In addition, its performance on the original benchmarks is also improved. Data and code are available at GitHub\footnote{\url{https://github.com/ruc-aimc-lab/nT2VR}}.
\end{itemize}
 


The remaining part is organized as follows. We discuss related work in Section \ref{sec:related}. The new evaluation protocol is described in Section \ref{sec:protocol}, followed by the NL method in Section \ref{sec:method} and experiments in Section \ref{sec:exp}. Major conclusions are presented in Section \ref{sec:concs}.

\section{Related Work} \label{sec:related}


\textbf{Progress on T2VR}.
Depending on whether visual / text encoders used to extract raw features from videos / queries are frozen, we divide current methods for T2VR into the following two groups, \ie feature re-learning methods \cite{icmr18-t2v,LiXirong2019W2VVPP,sigir20-t2v,dzabraev2021mdmmt,mm21-hanet} and end-to-end methods \cite{eccv2022-laff,luo2021clip4clip,fang2021clip2video}.

A feature re-learning method typically has a two-stage working pipeline. In the first stage, one or multiple pre-trained 2D/3D CNNs are used to extract frame-level or segment-level features from the video content, whilst pre-trained language models, \eg word2vec (w2v) or BERT, are adopted for extracting dense features from the text. Consider the W2VV++ series \cite{LiXirong2019W2VVPP,mm20-video-retrieval,LiXirong2020SEA} for instance. The method represents queries by concatenating the output of multiple existing text encoders. CE \cite{liu2019use}, MMT \cite{gabeur2020multi_MMT} and MDMMT \cite{dzabraev2021mdmmt} exploit multiple visual features that capture motion, appearance, face and OCR, respectively. We refer to \cite{ChenHu2021} for more details about the choice of the features. In the second stage, the pre-extracted visual / textual features are fed into a cross-modal representation learning network so that the re-learned features can be directly used for cross-modal matching. The choice of the network varies, raging from a simple feedforward network used by W2VV++, multi-level encoding networks used by DualEncoding \cite{tpami21-de}, DualTask \cite{Wu2020InterpretableEF}, HGR \cite{chen2020fine} and HANet \cite{mm21-hanet}, to more complicated graph auto-encoders used by FCA-Net \cite{mm21-finegrained}. The capability of these methods to handle negation is subject to their raw textual features. If these features are initially not discriminative to negation, their ability to represent negation is unlikely to be improved by feature re-learning.

Thanks to the advent of CLIP (Contrastive Language-Image Pre-Training) \cite{2021clip_icml}, end-to-end methods for T2VR have been developed recently. Built on the top of CLIP, CLIP-FT \cite{eccv2022-laff}, CLIP4Clip \cite{luo2021clip4clip} and CLIP2Video \cite{fang2021clip2video} have shown superior performance over the feature re-learning methods on multiple T2V benchmark datasets including MSR-VTT \cite{MSR-VTT}, MSVD \cite{msvd} and VATEX \cite{VATEX}. However, their ability to answer nT2VR is  unknown so far.

Wu and Ngo \cite{Wu2020InterpretableEF} describe briefly a boolean operation to tackle queries with negation. In that work, a total of five queries were manually created, \ie \emph{beach not man}, \emph{face not woman}, \emph{drinking not wine or beer}, \emph{flower not red or yellow}, and \emph{two people kissing not bride and groom}. Per query, \eg \emph{face not woman}, its score to a given video is computed as the video's similarity to the positive subquery (\emph{face}) subtracted by its similarity to the negative subquery (\emph{woman}). Note that the above operation is essentially post-processing, leaving the problem of negative learning untouched.

Earlier efforts have been made on exploiting negative feedback for T2VR, yet all in an \emph{interactive} search mode. For instance, Cooper \etal \cite{Fxpal2005} describe an interactive video retrieval system where a user can manually label the currently retrieved video shots either as positive or negative  (a.k.a. non-relevant). The system then exploits the negative shot set to implement negative reinforcement / feedback. As such, the user’s negative intent has to be specified after the first-round search and indirectly via labeling specific shots as negative. By contrast, this paper is targeted at \emph{automated} search, allowing a user to directly express what she or he does not want in a natural-language query in the first place. Hence, the proposed negation-aware video retrieval is conceptually novel and technically orthogonal to negative feedback.

\textbf{Understanding Negation in Large-scale Language Modeling}.
Large-scale pre-trained language models (PLM), as exemplified by BERT \cite{bert}, have demonstrated impressive performance on varied NLP tasks. However, recent works report that PLM's comprehension over negation is not satisfying \cite{kassner-schutze-2020-negated,whatbertisnot,BertUnderstandingNot}.  Kassner and Sch{\"u}tze show by their experiments that the probability for PLM to generate ``Birds cannot fly'' is nearly the same as ``Birds can fly'' \cite{kassner-schutze-2020-negated}. Hosseini \etal \cite{BertUnderstandingNot} report that when filling the blank with negation, \eg ``The macOS was not developed by \underline{\space\space\space}'', BERT answered with ``Apple'' in spite of the negator. To improve negation understanding of BERT, the authors propose to use an unlikelihood objective on negated sentences. Targeted at NLP tasks, the above technique is not directly applicable for addressing nT2VR.

\section{Proposed Evaluation Protocol} \label{sec:protocol}

As we have noted earlier, benchmark for evaluating nT2VR is nonexistent. We choose to re-purpose two public video-caption datasets, \ie MSR-VTT \cite{MSR-VTT} and VATEX \cite{VATEX}, commonly used in the T2VR literature. This is achieved by (partially) negating original queries 
and composing novel and controlled queries in Section \ref{ssec:new-query}.
Evaluation criteria suited for the re-purposed datasets are presented in Section \ref{ssec:criterion}. All this provides a new evaluation protocol for comprehensively assessing an (existing) T2VR model's ability to handle original, negated, and composed queries. 





\subsection{New Query Construction} \label{ssec:new-query}

\subsubsection{Negated Query Construction} \label{ssec:negated}

In order to construct (partially) negated queries from the original video captions, we use a simple rule-based strategy as follows. Given a caption $q$, we first use the NLTK\footnote{\url{https://github.com/nltk/nltk}} part-of-speech (POS) tagging API to identity verbs (VERB) and auxiliary verbs (AUX). Then, a negation cue is inserted right before the identified verb, \eg \emph{while} [not] \emph{dancing with many other people}, or after the AUX, \eg \emph{there is } [not] \emph{a fight at a basketball game}, yielding a negated variant of the caption, denoted by $q^-$. Due to the richness of the video content, a caption typically contains multiple verbs. In such cases, one of the verbs or AUX is randomly chosen to be negated, making $q^-$ partially negated \wrt $q$. See the appendix for more instances.


There also exists a relatively small amount of captions originally having negation cues, \eg \textit{a boy running is running without dress}. For these captions, we negate their original meaning by removing the negation cues, \eg \emph{a boy running is running} [with] \emph{dress}.

\begin{figure}[tb!]
\subfigure[Negated queries and videos known to be irrelevant]
{
\includegraphics[width=\columnwidth]{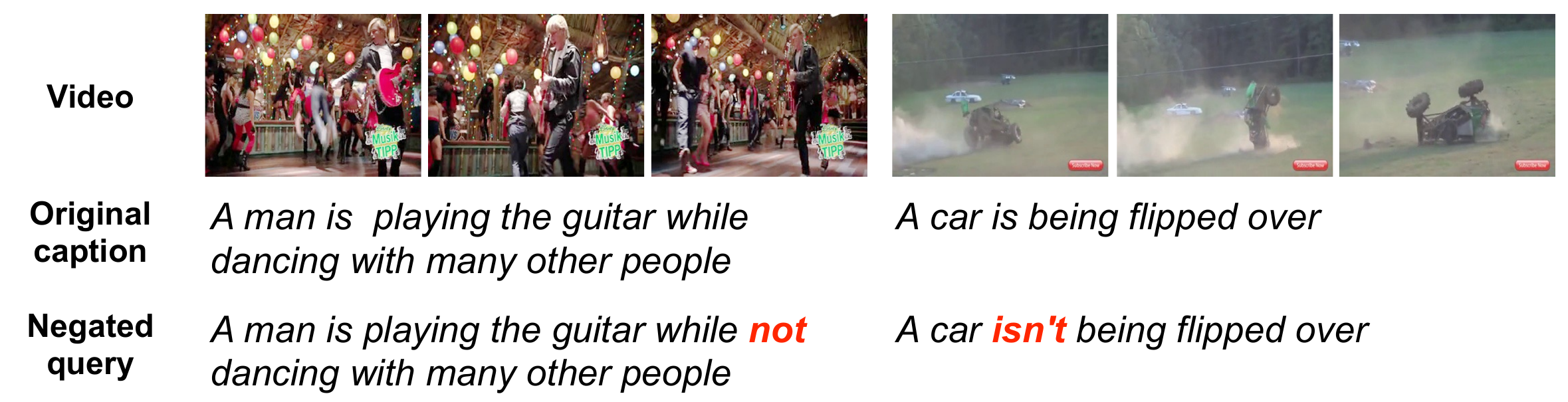}
}
\subfigure[Composed queries and videos known to be relevant]
{
\includegraphics[width=\columnwidth]{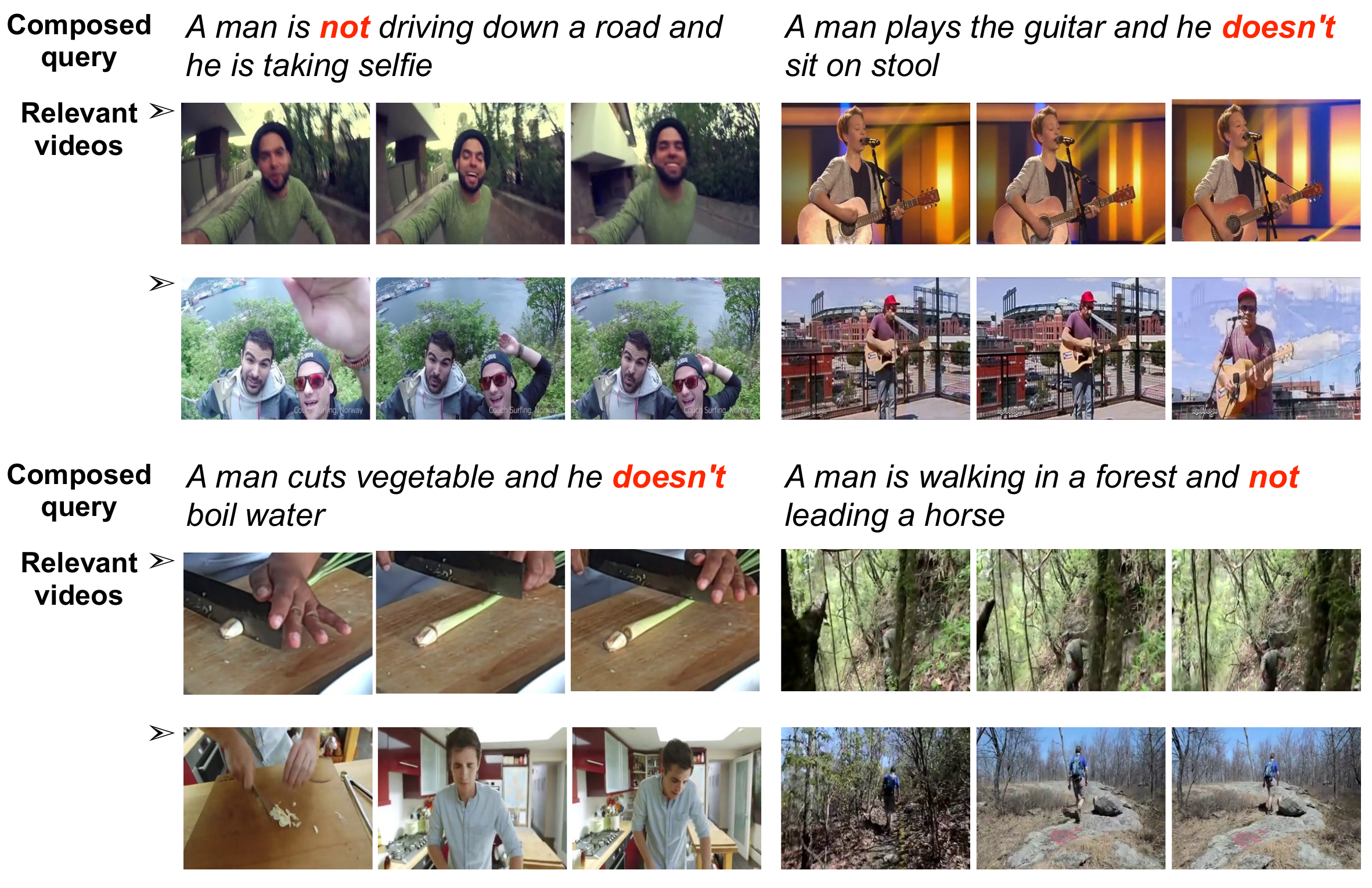}
}
\caption{Illustration of (a) negated and (b) composed queries in the proposed evaluation protocol. 
Data source: MSR-VTT. }
\label{fig:show-queries}
\end{figure}

When using $q^-$ as a query, the video corresponding to $q$ now becomes negative, see Fig. \ref{fig:show-queries}(a). Hence, a T2VR model that can handle negation shall rank the video lower. However, a downside of the negated query is that we are unsure which other videos are truly relevant \wrt the negated query. Evaluating on the negated queries alone is thus insufficient. Next, we propose to compose queries with negation that have reference videos available.



\subsubsection{Composed Query Construction} \label{ssec:composed}

For constructing composed queries, we first extract two linguistic groups, \ie subjects and verb phrases (VP), from the captions. Instances of subjects are \emph{a man}, \emph{people} and \emph{a car}, while instances of VPs are \emph{take selfie}, \emph{drive down a road} and \emph{play basketball}. 
Algorithm \ref{alg:text-parsing} shows python-style pseudo code for extracting  pairs of subjects and VPs from an (unrestricted) video caption.

\begin{algorithm}[h!]
	
  \caption{Subject and verb phrase (VP) extraction by NLTK chunk parsing} \label{alg:text-parsing}
   \textbf{Input}: A video caption \emph{q} \\
   \textbf{Output}: A list of (subject, VP) pairs \emph{s\_vp\_list} \\
  
  \begin{lstlisting}[language=python] 
        #Tag patterns for specific types of chunks
        grammar = '''
         NP: {<DT|JJ|NN.*>*<NN.*>} # A noun-phrase chunk
         PP: {<IN|RP><NP>} # A prepositional-phrase chunk
         VP: {<VB.*><NP|PP|CLAUSE>*} # A verb-phrase chunk
         CLAUSE: {<NP><VP>}  # A clause chunk
         '''
        chunk_parser = nltk.RegexpParser(grammar)
        tokens = nltk.word_tokenize(q)
        tagged_tokens = nltk.pos_tag(tokens)
        chunked_text = chunk_parser.parse(tagged_tokens)
        
        s_vp_list = []
        for vp in chunked_text.VPs:
            subject = find_subject(checked_text, vp)
            s_vp_list.append((subject, vp))
  \end{lstlisting}
\end{algorithm}

\begin{figure}[htbp!] 
\centering
\includegraphics[width=\columnwidth]{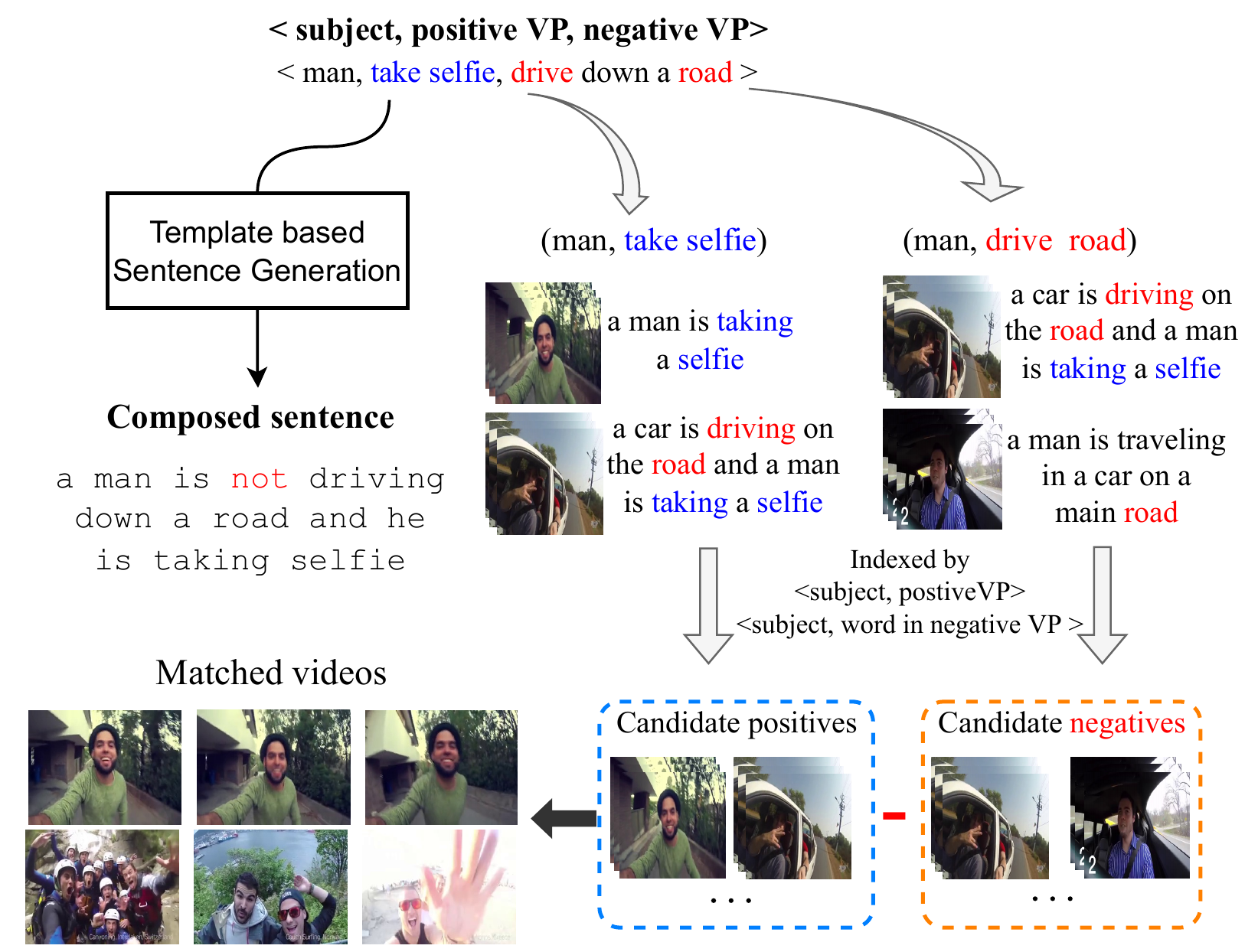}
\caption{\textbf{Key data flow of constructing a composed query and its matched videos}. Given a subject (\emph{man}), a positive VP (\emph{take selfie}) and a negative VP (\emph{drive down a road}), we use template-based sentence generation to obtain a composed query (\emph{a man is taking selfie and he is not driving down a road}). To find matched videos in the training data, we conduct phrase-level text retrieval on the video captions to identify a set of candidate positive videos and word-level text retrieval to identity a set of candidate negative videos. The matched videos are obtained by set difference.} 
\label{Figure:buildcomposed} 
\end{figure}

In order to ensure both linguistic and semantic soundness, our composed queries consist of a subject followed by two VPs, one used as positive, while the other used as negative. Given the above triplet, a novel query is produced by template-based sentence generation. For instance, given <\emph{man}, \emph{take selfie}, \emph{drive down a road}>, we have the following two queries: \emph{a man is taking selfie and he is not driving down a road} and \emph{a man is not driving down a road and he is taking selfie}. As illustrated in Fig. \ref{Figure:buildcomposed}, in order to find reference videos in a given training set, we use the positive VP to conduct phrase-level text retrieval on the video captions to identify a set of candidate positive videos. In order to exclude false positives, we favor precision over recall. Therefore, we use each word from the negative VP to perform word-level text retrieval to identity videos that are positive \wrt the word and thus being possibly negative \wrt the composed query. By a set-difference operation between the positive video set and the negative set, matched videos are found. It is possible that the operation may produce an empty set. In this case, the composed query will be discarded. By doing so, we effectively remove queries that describe scenes that are either counter-fact or rarely occur in the real world.

\subsubsection{Re-purposed Datasets} \label{ssec:new-datasets}

We perform negated / composed query construction on MSR-VTT and VATEX. For MSR-VTT, we adopt two data-split editions. One is provided by the dataset developers \cite{MSR-VTT}, with 3k test videos, while the other is specified by Yu \etal \cite{JSFusion} with 1k test videos. The two editions are referred to as MSR-VTT3k and MSR-VTT1k, respectively. For VATEX, we follow the data split\footnote{The number of videos used in this work is slightly less than the official number, \\as some videos were no longer available when downloading.} by Chen \etal \cite{chen2020fine}. The amount of original, negated and composed queries per test set is summarized in Table \ref{tab:datasets}.

\begin{table}
\renewcommand{\arraystretch}{1.1}
\caption{\textbf{Public datasets re-purposed by this paper for evaluating text-to-video retrieval (T2VR) with negation} (nT2VR).}
\label{tab:datasets}
\scalebox{0.95}{
\begin{tabular}{@{}lrrrr@{}}
\toprule 
\multicolumn{1}{@{}l}{\multirow{2}{*}{\textbf{Dataset}}} & \multicolumn{1}{c}{\multirow{2}{*}{ \textbf{Test videos}}} & \multicolumn{3}{c}{\textbf{Test queries}} \\ \cline{3-5} 
\multicolumn{1}{c}{} & \multicolumn{1}{c}{} & \emph{Original} & \emph{Negated} & \emph{Composed} \\ \hline
MSR-VTT3k \cite{MSR-VTT}  & 3,000 & 59,800 & 59,668 & 18,157 \\
MSR-VTT1k \cite{JSFusion}  & 1,000 & 1,000 & 923 & 3,697 \\
VATEX \cite{VATEX} & 1,398 & 13,980 & 7,339 & 8,394 \\
\bottomrule 
\end{tabular}
}

\end{table}


For a quick quality assessment, we randomly sampled 300 composed queries,  and manually checked whether the matched videos are truly relevant \wrt the queries. About 90\% of the queries are correctly associated with relevant videos. The composed queries are of sufficient quality for reliable evaluation.



\subsection{Evaluation Criteria} \label{ssec:criterion}

For the original test queries, we report the commonly used Recall at Rank $N$ ($R@N$, $N=1, 5, 10$), \ie the percentage of test queries that have their answers successfully retrieved among the top-$N$ ranked videos. In addition, we report Mean Inverted Rank (MIR), which reflects how the reference videos are positioned in the overall ranking list. The same metrics are used to evaluate a model's performance on the composed queries.


For each negated query $q^-$, we can trace back to its original query $q$, and thus know that the video associated with $q$ shall be negative \wrt $q^-$. Hence, the difference between the video's rank \wrt $q$ and that \wrt $q^-$ reflects a model's sensitivity to the introduced negation. In that regard, we report $\Delta R@N$, computed as
\begin{equation} \label{deltarank}
\Delta R@N(q^-) = R@N(q)- R@N(q^-). \\
\end{equation}
In a similar manner we compute $\Delta MIR$.

Ideally, a model more sensitive to negation shall produce larger values regarding both $\Delta R@N$ and $\Delta MIR$. However, using the $\Delta$ metrics alone is insufficient. Consider, for instance, a trivial solution that simply reverses the ranking list. Larger $\Delta R$  or $\Delta MIR$ shall not be interpreted as better retrieval performance.  Therefore, the performance on the composed query set shall be treated as primary, while the performance on the negated query set is secondary.






\section{Proposed Learning-based Method}  \label{sec:method}

In order to make our paper more self-contained, we first present some preliminaries concerning current deep learning based T2VR methods in Section \ref{ssec:pre-method}. How to handle negation in a learning-based manner is depicted in Section \ref{ssec:nl}.

\subsection{Preliminaries} \label{ssec:pre-method}

At a high level, a deep learning based T2VR model $\mathcal{M}$ works as follows. Given a textual query $q$ and a short video $x$, the model uses a text encoder, denoted by $\mathcal{M}_t$, and a visual encoder, denoted by $\mathcal{M}_v$, to project the query and the video into a $d$-dimensional common space. We use $\mathcal{M}_t(q)$ and $\mathcal{M}_v(x)$ to indicate the resultant textual and visual embedding vectors, respectively. A text-video similarity $s(x,q)$ is typically computed as the cosine similarity between the two embeddings \cite{tpami21-de,LiXirong2020SEA}. Accordingly, T2VR on a video collection is achieved by first scoring each video by $s(x,q)$ and then sorting them in descending order to acquire the top-ranked videos.

For model training, a set of manually captioned videos are required for jointly optimizing $\mathcal{M}_t$ and $\mathcal{M}_v$ so that a video $x^+$ relevant \wrt a given query shall have a larger similarity than an irrelevant video $x^-$, \ie $s(x^+,q)>s(x^-,q)$. Such a constraint is commonly implemented by minimizing a triplet ranking loss $\ell_{tri}$ with hard-negative mining \cite{VSE++,tpami21-de,mm21-consistency,mm21-finegrained,mm21-hanet}. Given a caption $q$ and a video $x^+$ it is describing, let $x^{\#}$ be the hardest negative video, practically selected from a given mini-batch. The loss $\ell_{tri}$ is calculated as
\begin{equation} \label{eq:base-loss}
\left\{ 
\begin{array}{ll}
x^{\#} &=\argmax_{x^-} (s(x^-,q) - s(x^+,q)) \\ 
\ell_{tri}(q,x^{+}, x^{\#})  &=\max (0, m_0 +s(x^{\#},q) - s(x^+, q)),
\end{array} \right.
\end{equation}
where 
$m_0$ is a positive hyper-parameter controlling the margin. 

\subsection{Negation Learning} \label{ssec:nl}

Given $(x^+,q)$ as paired video and caption, we can effectively augment the training data by using the negator described in Section \ref{ssec:negated} to generate a soft-negative caption $q^-$ \wrt the video $x$. Intuitively, we shall have $sim(x^+, q) > sim(x, q^-)$. So a relatively straightforward strategy to perform negation learning (NL) is to compute another $\ell_{tri}$ for the triplet $<x^+, q, q^->$,
\begin{equation} \label{eq:simple-nl}
\ell_{tri}(x^+, q, q^-)  = \max (0, m_1 +s(x^+,q^-) - s(x^+, q)),
\end{equation}
where $m_1$ is a margin parameter. Adding this auxiliary loss to the primary loss in Eq. \ref{eq:base-loss}, we obtain a joint loss $\ell_{snl}(q,x^+)$  as
\begin{equation} \label{eq:snl}
\ell_{snl}(q,x^+) = \ell_{tri}(q,x^{+}, x^{\#}) + \lambda_1  \ell_{tri}(x^+, q, q^-),   
\end{equation}
with $\lambda_1$ as a weight. The model $\mathcal{M}$ is trained by minimizing $\ell_{snl}(q,x^+)$. We term this strategy \emph{Simple Negation Learning} (SNL).

We see from Eq. \ref{eq:simple-nl} that SNL treats $q^-$ as a common negative caption \wrt the video. Recall that $q^-$ is derived from $q$ by negating one of its clauses. The unchanged part inherited from $q$, \eg \emph{a man is playing the guitar} as illustrated in Fig. \ref{fig:show-queries}, remains semantically relevant to the video content. This means the negative pair $(x^+, q^-)$ shall maintain certain similarity. In other words, when viewing $x^+$ as a pivot point in the common space where $q$ shall be more close to $x^+$, $q^-$ shall not be pushed too far away from $x^+$. To that end, while $s(x^+,q)$ shall be larger than $s(x^+,q^-)$, there needs to be an upper boundary on their difference, \ie  $s(x^+,q) - s(x^+,q^-) < m_2$, with $m_1 < m_2 < 2$. We modify Eq. \ref{eq:simple-nl} to take the new constraint into account, resulting in a \emph{bidirectionally} constrained loss $\ell_{bcl}$ as
\begin{equation} \label{eq:bcl}
\begin{array}{ll}
    \ell_{bcl}(x^+,q,q^-) = & \max (0, m_1 +s(x^+,q^-) - s(x^+, q)) + \\
    & \max (0, - m_2 - s(x^+,q^-) + s(x^+, q)).
\end{array}
\end{equation}

Similarly, given the original caption $q$, we expect that its cross-modal similarity to its relevant video $x^+$ shall be larger than its uni-modal similarity to its negated variant $q^-$. Meanwhile, the gap between $s(x^+,q)$ and $s(q,q^-)$ shall be bounded. To that end, we compute $\ell_{bcl}$ for the triplet $<q, x^+, q^->$ as
\begin{equation} \label{eq:bcl2}
\begin{array}{ll}
    \ell_{bcl}(q,x^+,q^-) = & \max (0, m_3 +s(q,q^-) - s(q, x^+)) + \\
    & \max (0, - m_4 - s(q, q^-) + s( q,x^+)),
\end{array}
\end{equation}
where $m_3$ and $m_4$ are margin parameters with $0<m_3<m_4<2$.

Note that $\ell_{bcl}(x^+,q,q^-)$ and $\ell_{bcl}(q,x^+,q^-)$ respectively use the video $x^+$ and the original caption $q$ as a pivot in the common space to exploit the negated information. 
By jointly minimizing the two losses, the model $\mathcal{M}$ is trained to find a proper embedding for the soft negative $q^-$ \wrt both the video $x$ and the original caption $q$. Accordingly, we term the improved strategy \emph{Bidirectional Negation Learning} (BNL), with the corresponding loss defined as 
\begin{equation} \label{eq:bnl}
\ell_{bnl}(q,x^+) = \ell_{tri}(q,x^{+}, x^{\#}) + \lambda_2 (\ell_{bcl}(x^+,q,q^-) + \ell_{bcl}(q,x^+,q^-)), \end{equation}
where $\lambda_2$ is a small positive weight for balancing the primary and the auxiliary losses.

\subsection{Choice of the T2VR Model} 

We instantiate $\mathcal{M}$ with CLIP (ViT-B/32) \cite{2021clip_icml}. Originally developed for text-image matching, CLIP consists of a BERT for text embedding and a Vision Transformer (ViT) for image embedding. To deal with the video input, we use ViT to extract features per frame, and aggregate the frame-level features to the video level by mean pooling\footnote{Mean pooling can be replaced by attention-based pooling for better performance~\cite{mm21-mmmil}.} for subsequent cross-modal similarity learning and matching. We use CLIP-\emph{snl} and CLIP-\emph{bnl} to indicate CLIP trained with $\ell_{snl}$ and $\ell_{bnl}$, respectively.

Note that our NL methods are model-agnostic, so other end-to-end alternatives to CLIP can in principle be used. We leave this for future exploration.

\section{Experiments} \label{sec:exp}

  \begin{table*}[tbh!]
\renewcommand{\arraystretch}{1.1}
\caption{\textbf{Performance on the original, negated and composed query sets of the re-purposed MSR-VTT3k}. 
The boolean operation is not applicable to the original queries. Our CLIP-\emph{bnl} tops the performance on the composed query set, while being sensitive on the negated query set.}
\begin{tabular}{@{}lrrrrrrrrrrrrrrrr@{}}
\toprule 
\multicolumn{1}{@{}l}{\multirow{2}{*}{\textbf{Models}}} &  & \multicolumn{4}{c}{\textbf{Original} ($\uparrow$)}  &  & \multicolumn{4}{c}{\textbf{Negated} ($\uparrow$)} &  & \multicolumn{4}{c}{\textbf{Composed} ($\uparrow$)} \\ \cline{3-6} \cline{8-11} \cline{13-16} 
 &  & \multicolumn{1}{l}{$R1$} & \multicolumn{1}{l}{$R5$} & \multicolumn{1}{l}{$R10$} & \multicolumn{1}{l}{$MIR$} &  & \multicolumn{1}{l}{$\Delta R1$} & \multicolumn{1}{l}{$\Delta R5$} & \multicolumn{1}{l}{$\Delta R10$} & \multicolumn{1}{l}{$\Delta MIR$} &  & \multicolumn{1}{l}{$R1$} & \multicolumn{1}{l}{$R5$} & \multicolumn{1}{l}{$R10$} & \multicolumn{1}{l}{$MIR$} \\ \hline
W2VV++ \cite{LiXirong2019W2VVPP} &  & 11.4 & 29.9 & 40.7 & 0.208 &  &0.3 & 0.4 & 0.3 & 0.003 &  & 6.6 & 23.0 & 33.6 & 0.154 \\
SEA \cite{LiXirong2020SEA} &  & 12.4 & 32.0 & 43.4 & 0.224 &  & 0.1 & 0.3 & 0.3 & 0.002 &  & 7.5 & 24.3 & 34.9 & 0.164 \\
CLIP \cite{2021clip_icml} &  & 21.2 & 40.8 & 50.2 & 0.309 &  & 1.5 & 2.5 & 2.9 & 0.020 &  & 6.9 & 24.2 & 35.6 & 0.160 \\
CLIP* (\emph{this paper}) &  & 27.7 & 53.0 & 64.2 & 0.398 &  & 0.5 & 1.1 & 1.1 & 0.008 &  & 11.4 & 33.3 & 46.2 & 0.225 \\
CLIP4Clip \cite{luo2021clip4clip} &  & \textbf{28.9} & \textbf{54.4} & \textbf{65.1} & \textbf{0.410} &  & 0.8 & 1.5 & 1.2 & 0.010 &  & 11.3 & 33.3 & 45.6 & 0.222 \\
\hline
\textit{Boolean operation:} &  &  &  &  &  &  &  &  &  &  &  &  &  &  &  \\
W2VV++ &  & -- & -- & -- & -- &  & 10.5 & 26.1 & 34.6 & 0.182 &  & 8.9 & 23.7 & 32.5 & 0.166 \\
SEA &  & -- & -- & -- & -- &  & 11.9 & 29.1 & 38.2 & 0.202 &  & 7.5 & 19.8 & 27.9 & 0.142 \\
CLIP &  & -- & -- & -- & -- &  & 18.8 & 37.5 & 46.2 & 0.278 &  & 5.9 & 16.7 & 23.9 & 0.118 \\
CLIP* &  & -- & -- & -- & -- &  & 25.3 & 47.1 & 56.1 & 0.353 &  & 13.5 & 33.7 & 45.5 & 0.236 \\
CLIP4Clip &  & -- & -- & -- & -- &  &\textbf{27.2} &\textbf{ 51.0} & \textbf{59.9} & \textbf{0.380} &  & 8.0 & 22.9 & 32.0 & 0.158 \\ \hline
 CLIP-\emph{bnl}  (\emph{this paper}) && 28.4 & 53.7 & 64.6 & 0.404 &&  5.0 & 6.9 & 6.9 & 0.057 &  & \textbf{15.3} & \textbf{40.0} & \textbf{53.3} & \textbf{0.274}\\
\bottomrule 
\end{tabular}
\label{tab:eval_msrvtt3k}
\end{table*}

 \begin{table*}[tbh!]
\renewcommand{\arraystretch}{1.1}
\caption{\textbf{Performance on the re-purposed MSR-VTT1k}.}
\begin{tabular}{@{}lrrrrrrrrrrrrrrrr@{}}
\toprule 
\multicolumn{1}{@{}l}{\multirow{2}{*}{\textbf{Models}}} &  & \multicolumn{4}{c}{\textbf{Original} ($\uparrow$)} &  & \multicolumn{4}{c}{\textbf{Negated} ($\uparrow$)} &  & \multicolumn{4}{c}{\textbf{Composed} ($\uparrow$)} \\ \cline{3-6} \cline{8-11} \cline{13-16} 
 &  & \multicolumn{1}{l}{$R1$} & \multicolumn{1}{l}{$R5$} & \multicolumn{1}{l}{$R10$} & \multicolumn{1}{l}{$MIR$} &  & \multicolumn{1}{l}{$\Delta R1$} & \multicolumn{1}{l}{$\Delta R5$} & \multicolumn{1}{l}{$\Delta R10$} & \multicolumn{1}{l}{$\Delta MIR$} &  & \multicolumn{1}{l}{$R1$} & \multicolumn{1}{l}{$R5$} & \multicolumn{1}{l}{$R10$} & \multicolumn{1}{r}{$MIR$} \\ \hline
W2VV++ &  & 24.7 & 50.4 & 62.2 & 0.371 &  & 1.2 & -0.5 & 0.8 & 0.007  &  & 10.7 & 32.9 & 46.0 & 0.218 \\
SEA &  & 27.2 & 54.3 & 65.8 & 0.398 &  & -0.5 & -0.9 & -1.6 & -0.007 &  & 12.2 & 34.6 & 47.0 & 0.232 \\
CLIP &  & 31.6 & 54.2 & 64.2 & 0.422 &  & 1.4 & 1.4 & 1.5 & 0.017 &  & 12.9 & 35.0 & 46.2 & 0.237 \\
CLIP* &  & 41.1 & 69.8 & 79.9 & 0.543 &  & 0.0 & 1.7 & 1.0 & 0.006 &  & 17.3 & 46.8 & 61.2 & 0.310 \\
CLIP4Clip &  & \textbf{43.9} & \textbf{70.6} & \textbf{80.2} & \textbf{0.560 }&  & 1.2 & -1.7 & 0.0 & 0.008 &  & 15.0 & 43.1 & 57.8 & 0.281 \\
\hline
\textit{Boolean operation:} &  &  &  &  &  &  &  &  &  &  &  &  &  &  &  \\
W2VV++ &  & -- & -- & -- & -- &  & 21.3 & 40.9 & 51.2 & 0.310 &  & 11.2 & 27.9 & 36.9 & 0.196 \\
SEA &  & -- & -- & -- & -- &  & 23.9 & 47.6 & 54.1 & 0.344 &  & 10.9 & 26.6 & 35.6 & 0.188 \\
CLIP &  & -- & -- & -- & -- &  & 26.4 & 46.2 & 56.8 & 0.354 &  & 6.3 & 18.4 & 25.9 & 0.129 \\
CLIP* &  & -- & -- & -- & -- &  & 35.9 & 59.5 & 65.2 & 0.463 &  & 17.6 & 42.0 & 52.0 & 0.291 \\
CLIP4Clip &  & -- & -- & -- & -- &  & \textbf{40.0} & \textbf{61.9} & \textbf{69.1} & \textbf{0.495} &  & 8.5 & 25.6 & 34.9 & 0.171 \\
\hline
 CLIP-\emph{bnl} && 42.1 & 68.4 & 79.6 & 0.546 &  & 12.2 & 11.7 & 14.4 & 0.121 & & \textbf{24.8} & \textbf{57.6} & \textbf{68.8} & \textbf{0.391}\\
\bottomrule
\end{tabular}
\label{tab:eval_msrvtt1k}
\end{table*}

    \begin{table*}[tbh!]
\renewcommand{\arraystretch}{1.1}
\caption{\textbf{Performance on the re-purposed VATEX}.}
\begin{tabular}{@{}lrrrrrrrrrrrrrrrr@{}}
\toprule 
\multicolumn{1}{@{}l}{\multirow{2}{*}{\textbf{Models}}} &  & \multicolumn{4}{c}{\textbf{Original} ($\uparrow$)} &  & \multicolumn{4}{c}{\textbf{Negated} ($\uparrow$)} &  & \multicolumn{4}{c}{\textbf{Composed} ($\uparrow$)} \\ \cline{3-6} \cline{8-11} \cline{13-16} 
 &  & \multicolumn{1}{l}{$R1$} & \multicolumn{1}{l}{$R5$} & \multicolumn{1}{l}{$R10$} & \multicolumn{1}{l}{$MIR$} &  & \multicolumn{1}{l}{$\Delta R1$} & \multicolumn{1}{l}{$\Delta R5$} & \multicolumn{1}{l}{$\Delta R10$} & \multicolumn{1}{l}{$\Delta MIR$} &  & \multicolumn{1}{l}{$R1$} & \multicolumn{1}{l}{$R5$} & \multicolumn{1}{l}{$R10$} & \multicolumn{1}{l}{$MIR$}  \\ \hline
W2VV++ &  & 40.5 & 76.2 & 84.6 & 0.561 &  & 0.4 & 0.9 & 0.1 & 0.003 &  & 12.4 & 33.7 & 46.2 & 0.233 \\
SEA &  & 41.8 & 78.5 & 87.0 & 0.578 &  & -0.3 & 1.0 & 0.3 & 0.000 &  & 12.5 & 34.3 & 47.5 & 0.238 \\
CLIP &  & 41.4 & 72.9 & 82.7 & 0.555 &  & 1.9 & 2.1 & 2.2 & 0.018 &  & 10.5 & 28.3 & 41.3 & 0.201 \\
CLIP* &  & 56.8 & 88.4 & \textbf{ 94.4} & 0.703 &  & 0.2 & 0.4 & 0.7 & 0.004 &  & 14.2 & 39.2 & 53.3 & 0.266 \\
CLIP4Clip &  &\textbf{ 61.5 }& \textbf{88.8} & 94.0 &\textbf{ 0.734} &  &  0.8 & 0.3 & 0.6 & 0.006  &  & 14.3 & 38.4 & 51.5 & 0.263 \\
\hline
\textit{Boolean operation:} &  &  &  &  &  &  &  &  &  &  &  &  &  &  &  \\
W2VV++ &  & -- & -- & -- & -- &  & 31.5 & 57.1 & 61.6 & 0.421  &  & 11.6 & 31.6 & 42.1 & 0.215 \\
SEA &  & -- & -- & -- & -- &  &33.1 & 60.3 & 65.2 & 0.446 &  & 12.0 & 29.7 & 39.5 & 0.209 \\
CLIP &  & -- & -- & -- & -- &  & 32.5 & 57.2 & 64.5 & 0.431 &  & 5.0 & 18.0 & 25.6 & 0.116 \\
CLIP* &  & -- & -- & -- & -- &  & 45.7 & 69.6 & 71.9 & 0.554 &  & 14.1 & 34.4 & 45.1 & 0.243 \\
CLIP4Clip &  & -- & -- & -- & -- &  & \textbf{52.6 }& \textbf{72.7} & \textbf{75.8} & \textbf{0.609} &  & 8.7 & 25.8 & 34.3 & 0.171 \\
\hline
 CLIP-\emph{bnl} &  & 57.6 & 88.3 & 94.0 & 0.708 &  &14.0 & 11.7 & 8.6 & 0.125 &  & \textbf{16.6 }& \textbf{39.9} & \textbf{53.9} & \textbf{0.284} \\ 
\bottomrule 
\end{tabular}

\label{tab:eval_vatex}
\end{table*}



    
\subsection{Implementation Details}

Hyperparameters used in this work are empirically set as follows and fixed throughout our experiments, unless stated otherwise. 
The margin parameter $m_{0}$ for the primary retrieval loss (Eq. \ref{eq:base-loss}) is set to 0.2 according to VSE++~\cite{VSE++}. The lower and upper boundaries, \ie $m_1$ and $m_2$ for $\ell_{bcl}(x^+, q, q^-)$ (Eq. \ref{eq:bcl}) are set to 0.1 and 0.6, while $m_3$ and $m_4$ for $\ell_{bcl}(q,x^+,q^-)$ (Eq. \ref{eq:bcl2}) are set to $0.1$ and $0.3$. The weight $\lambda_1$ for SNL and $\lambda_2$ for BNL are both set to 1e-3. 


Our deep learning environment is PyTorch (1.7.0) \cite{PyTorch19} plus NVIDIA GEFORCE RTX 3090 GPUs. We perform SGD based training, with RMSProp as the optimizer. The learning rate is initially 1e-6, decayed by a factor of 0.99 per epoch. We use an early stopping strategy which stops training when no validation performance increase is achieved in two consecutive epochs. 



\subsection{Evaluating Current T2VR Models} \label{ssec:eval-tv2}



\subsubsection{T2VR Model Selection}

We choose the following models that have PyTorch training code publicly available: \\
$\bullet$ W2VV++\footnote{\href{https://github.com/li-xirong/w2vvpp}{https://github.com/li-xirong/w2vvpp}}, MM19 \cite{LiXirong2019W2VVPP}: This model learns to project text and video into a latent space, by using bag-of-words (bow), w2v and GRU as its text encoders and ResNeXt-101 / ResNet-152 pre-trained on ImageNet as its visual encoders. \\
$\bullet$  SEA\footnote{\href{https://github.com/li-xirong/sea}{https://github.com/li-xirong/sea}}, TMM21 \cite{LiXirong2020SEA}: SEA exploits four text encoders (bow, w2v, GRU, BERT) in a multi-space similarity learning framework. Its visual encoders are the same as W2VV++. \\
$\bullet$  CLIP\footnote{\href{https://github.com/openai/CLIP}{https://github.com/openai/CLIP}}, ICML21 \cite{2021clip_icml}: A text-image matching model pre-trained on 400 million image-text pairs collected from the Internet.\\
$\bullet$   CLIP*: We fine-tune the pre-trained CLIP on each of the three training sets (Table \ref{tab:datasets}), using the retrieval loss as expressed in Eq. \ref{eq:base-loss}. \\
$\bullet$  CLIP4Clip\footnote{\href{https://github.com/ArrowLuo/CLIP4Clip}{https://github.com/ArrowLuo/CLIP4Clip}}, arxiv21 \cite{luo2021clip4clip}:  An end-to-end model which transfers the knowledge of the CLIP model for T2VR. 

Among them, W2VV++ and SEA are feature re-learning based, while the others are all end-to-end. 




 \subsubsection{Results}

 The performance of the selected T2VR models on the re-purposed MSR-VTT3k, MSR-VTT1k and VATEX is shown in Table \ref{tab:eval_msrvtt3k}, \ref{tab:eval_msrvtt1k}, and \ref{tab:eval_vatex}, respectively.  The CLIP series (CLIP / CLIP* / CLIP4Clip), due to their end-to-end learning ability, clearly outperform the two feature re-learning alternatives (W2VV++ / SEA) on the original query set.
 However, the performance difference between the CLIP series  and feature re-learning on the negated query set is much smaller, suggesting that they are insensitive to the negation. Note that the pre-trained CLIP has the relatively largest $\Delta MIR$ of 0.020, see Row\#3 in Table \ref{tab:eval_msrvtt3k}. Similar results can also be observed from Table  \ref{tab:eval_msrvtt1k} and \ref{tab:eval_vatex}. Recall that different from CLIP, all the other models have been re-trained on the video-description data. These results are consistent with our earlier observation that the video descriptions lack negation. Learning from such data makes the models even more insensitive to negation in queries.

 Our CLIP-\emph{bnl} performs relatively lower than the SOTA model (CLIP4Clip) on the original query set. Note that we have no intention to beat the SOTA. The inclusion of the SOTA in our experiments is mainly to answer the question raised in the beginning, \ie, \emph{do the (current) models understand what the users do not want}? The experiments show that despite its leading performance on the original queries, CLIP4Clip is unaware of negation in queries, as demonstrated by its small  and  on the negated query set and its lower R1/R5/R10/MIR scores on the composed query set. Consider the MIR metric for instance, CLIP-\emph{bnl} outperforms CLIP4Clip with a clear margin: 0.222 $\rightarrow$  0.274 on MSR-VTT3k (23.4\% relative improvement), 0.281 $\rightarrow$ 0.391 on MSR-VTT1k (39.1\%) and 0.263 $\rightarrow$ 0.284 on VATEX (8.0\%). Such performance gaps are deemed to be significant in the literature of T2VR.

\subsection{NL versus Alternatives} \label{ssec:eval-nl-vs-others}

\subsubsection{Baselines} 

We implement the boolean operation  \cite{Wu2020InterpretableEF} for each of the T2VR models previously evaluated. The operation requires decomposing a given query into a positive subquery and a negative subquery. Handling a composed query is relatively straightforward, as we know which part of the query is positive and which is negative. For a negated query, although the negation cue is known, the negation scope followed by the cue is unknown. We resort to negBERT \cite{Khandelwal2020NegBERTAT} to automatically detect the negation scope. The detected result is used as the negative subquery, while the remaining part of the query is used as the positive subquery.

\subsubsection{Results}

As shown in Table \ref{tab:eval_msrvtt3k}, \ref{tab:eval_msrvtt1k} and \ref{tab:eval_vatex}, the T2VR models with the boolean operation produces much larger response on the negated query set, when compared to their counterparts w/o the operation. This indicates that the boolean operation makes the models more sensitive to negation. However, the higher sensitivity is obtained at the cost of the undesired performance drop on the composed query set. Consider CLIP4Clip, the top performer on the original query set for instance. With the boolean operation, its MIR score on the composed query decreases: 0.222 $\rightarrow$ 0.158 on MSR-VTT3k, 0.281 $\rightarrow$ 0.171 on MSR-VTT1k, and 0.263 $\rightarrow$ 0.171 on VATEX. We conclude that the boolean operation is not effective for dealing with the negation in composed queries. By contrast, our CLIP-\emph{bnl}, obtained by fine-tuning CLIP with the proposed bidirectional negation learning, shows superior performance on the composed query set. 
In addition, CLIP-\emph{bnl} also exhibits higher sensitivity on the negative set against its NL-free counterpart, \ie CLIP* ($\Delta MIR$ 0.057 versus 0.008, 0.121 versus  0.006, and 0.125 versus 0.004 on MSR-VTT3k, MSRV-VTT1k, and VATEX, respectively).

\subsection{Ablation Study on NL} \label{ssec:eval-ab}

In order to verify the necessity of BNL against SNL, we try with varied implementation choices of the auxiliary loss.  Per choice, the related margin parameters are set based on the performance on the held-out validation set. The ablation study is conducted on MSR-VTT3k. 

\begin{table*}[tbh!]
\renewcommand{\arraystretch}{1.1}
\centering
\begin{center}
\caption{\textbf{The influence of the auxiliary loss}. Each row is the performance of a specific CLIP model trained by weightedly adding the corresponding auxiliary loss to the primary retrieval loss. Dataset: MSR-VTT3k. }
\label{tab:BNL}
\scalebox{1}{
\begin{tabular}{@{}lrrrrlrrrrlrrrr@{}}
\toprule
\multirow{2}{*}{\textbf{Auxiliary loss}} & \multicolumn{4}{c}{\textbf{Original}  ($\uparrow$)} &  & \multicolumn{4}{c}{\textbf{Negated} ($\uparrow$)} & \multicolumn{1}{c}{} & \multicolumn{4}{c}{\textbf{Composed}  ($\uparrow$)} \\ \cline{2-5} \cline{7-10} \cline{12-15} 
 & \multicolumn{1}{l}{$R1$} & \multicolumn{1}{l}{$R5$} & \multicolumn{1}{l}{$R10$} & \multicolumn{1}{l}{$MIR$} &  & \multicolumn{1}{l}{$\Delta R1$} & \multicolumn{1}{l}{$\Delta R5$} & \multicolumn{1}{l}{$\Delta R10$} & \multicolumn{1}{l}{$\Delta MIR$} &  & \multicolumn{1}{l}{$R1$} & \multicolumn{1}{l}{$R5$} & \multicolumn{1}{l}{$R10$} & \multicolumn{1}{l}{$MIR$} \\
  \hline
 -- (Row\#4, Table \ref{tab:eval_msrvtt3k}) & 27.7 & 53.0 & 64.2 & 0.398 & \multicolumn{1}{r}{} & 0.5 & 1.1 & 1.1 & 0.008 & \multicolumn{1}{r}{} & 11.4 & 33.3 & 46.2 & 0.225 \\ 
 \hline
\multicolumn{3}{@{}l}{\textit{Simple Negation Learning (SNL):}} & \multicolumn{1}{l}{} & \multicolumn{1}{l}{} &  & \multicolumn{1}{l}{} & \multicolumn{1}{l}{} & \multicolumn{1}{l}{} & \multicolumn{1}{l}{} &  & \multicolumn{1}{l}{} & \multicolumn{1}{l}{} & \multicolumn{1}{l}{} & \multicolumn{1}{l}{} \\
$\ell_{tri}(x^+, q, q^-)$ & \textbf{28.8} & \textbf{54.2} & \textbf{65.1} & \textbf{0.408} &  & 1.8 & 2.5 & 2.5 & 0.021 &  & 12.9 & 35.1 & 48.1 & 0.241 \\
$\ell_{tri}(q, x^+, q^-)$ & 28.1 & 53.5 & 64.6 & 0.402 &  & \textbf{21.4} & \textbf{38.8} & \textbf{44.5} & \textbf{0.288} &  & 7.4 & 19.4 & 27.4 & 0.141 \\ \hline
\multicolumn{3}{@{}l}{\textit{Bidirectional Negation Learning (BNL):}} & \multicolumn{1}{l}{} & \multicolumn{1}{l}{} &  & \multicolumn{1}{l}{} & \multicolumn{1}{l}{} & \multicolumn{1}{l}{} & \multicolumn{1}{l}{} &  & \multicolumn{1}{l}{} & \multicolumn{1}{l}{} & \multicolumn{1}{l}{} & \multicolumn{1}{l}{} \\

$\ell_{bcl}(x^+, q, q^-)$ & 27.5 & 52.7 & 63.8 & 0.395 &  & 2.7 & 3.7 & 3.6 & 0.030 &  & 14.6 & 38.9 & 51.5 & 0.264 \\
$\ell_{bcl}(q, x^+, q^-)$ & 28.0 & 53.3 & 64.2 & 0.400 &  & 15.8 & 26.0 & 28.1 & 0.199 &  & 14.8 & 36.1 & 50.4 & 0.257 \\
$\ell_{bcl}(x^+, q, q^-)+\ell_{bcl}(q, x^+, q^-)$ & 28.4 & 53.7 & 64.6 & 0.404 &  & 5.0 & 6.9 & 6.9 & 0.057 &  & \textbf{15.3} & \textbf{40.1} & \textbf{53.3} & \textbf{0.274} \\
\bottomrule
\end{tabular}
}
\end{center}

\end{table*}

\begin{figure*}
    \subfigure[Original query set] 
    {\includegraphics[width=0.6\columnwidth]{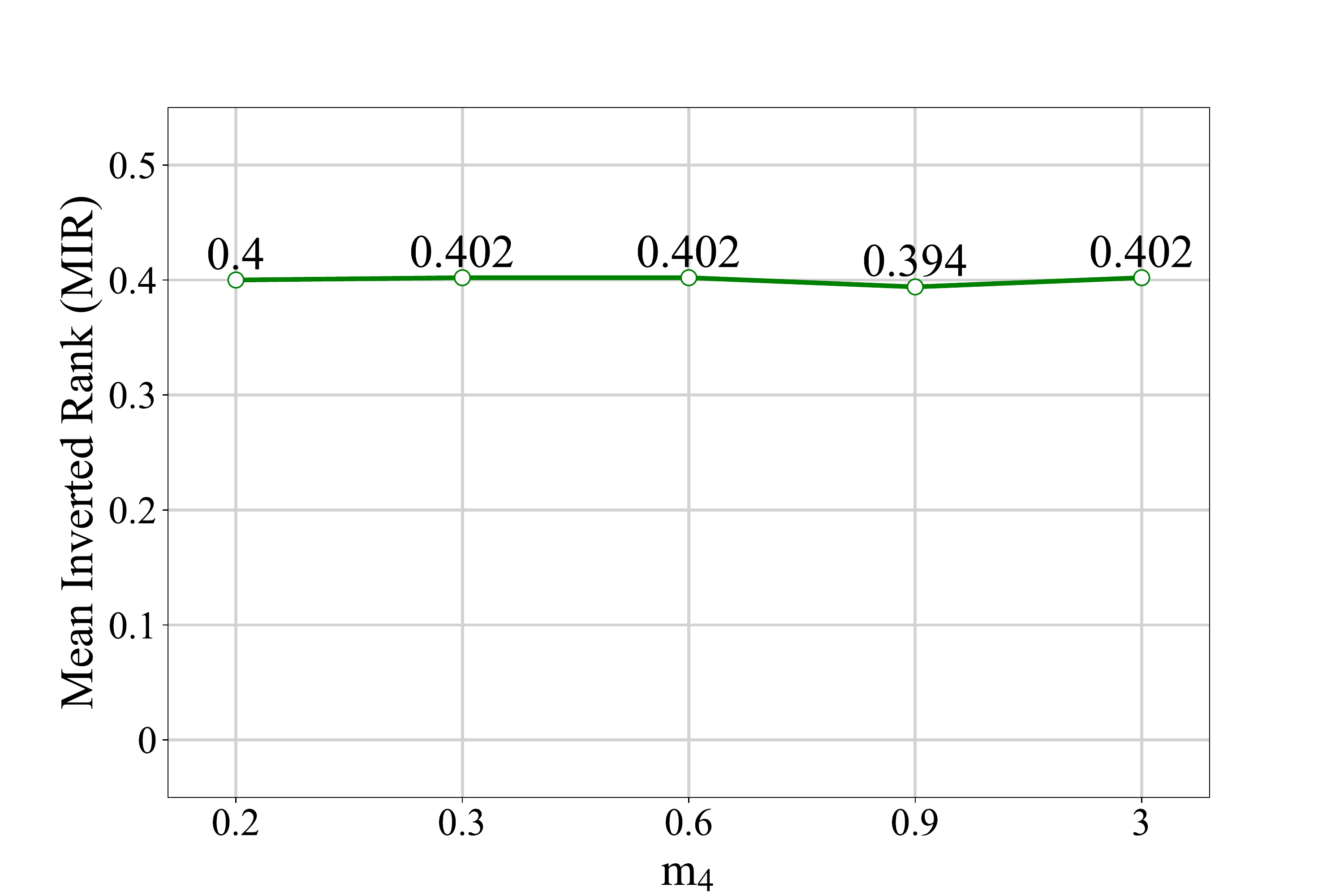}} 
    \subfigure[Negated query set] 
    {\includegraphics[width=0.6\columnwidth]{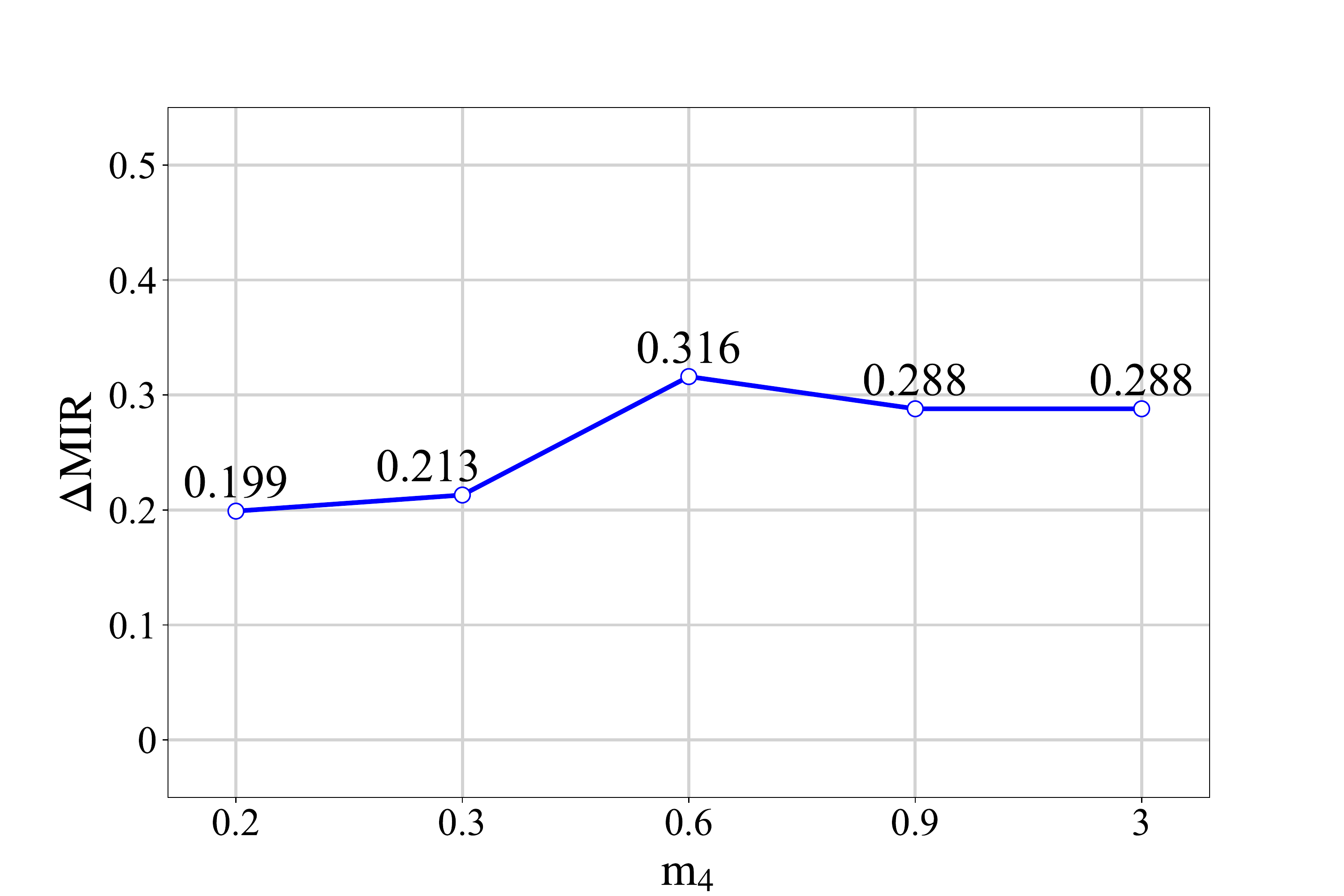}} 
    \subfigure[Composed query set] 
    {\includegraphics[width=0.6\columnwidth]{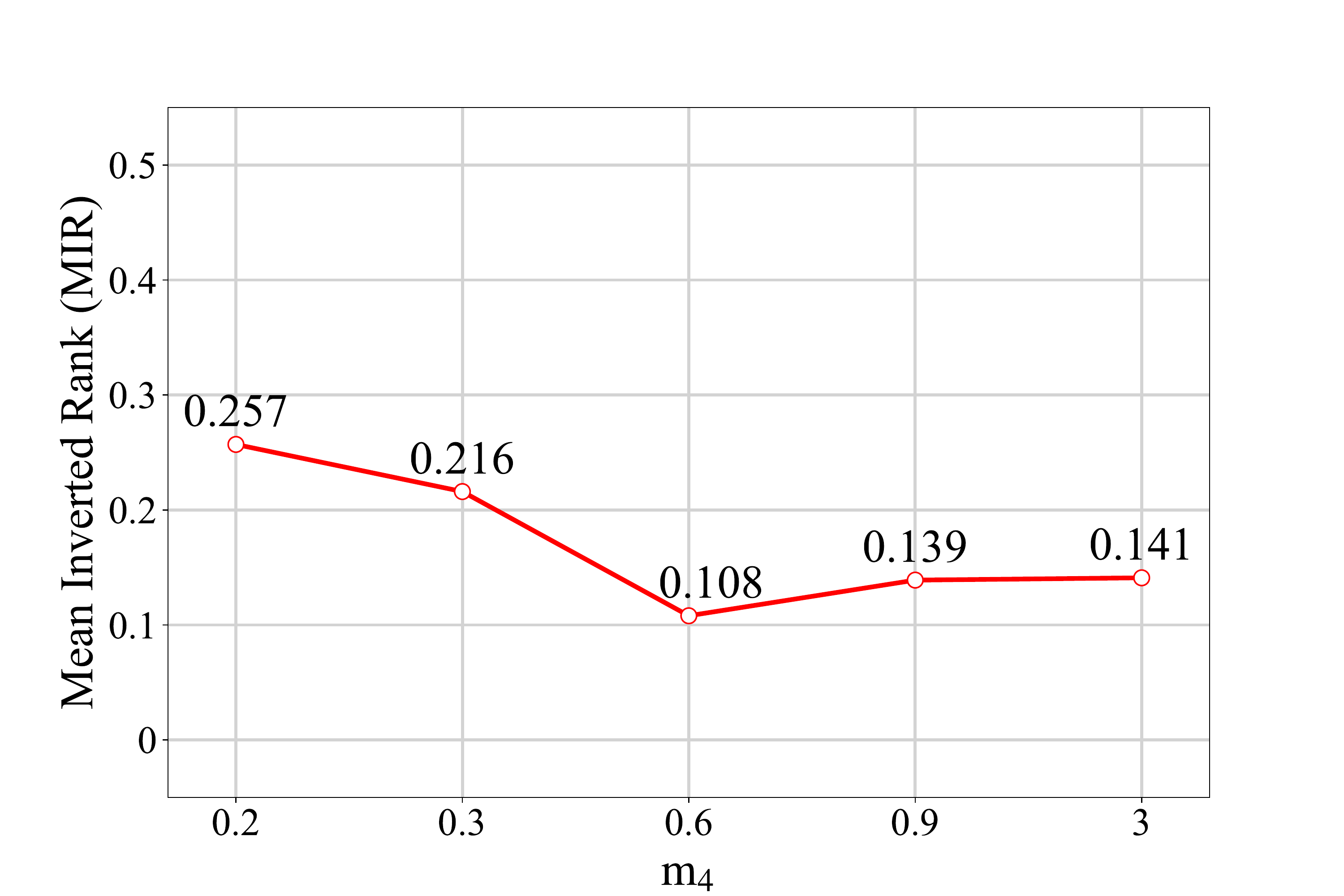}} 
    \caption{\textbf{Performance curves \wrt the upper boundary $m_4$ in $\ell_{bcl}(q, x^+, q^-)$}. The auxiliary loss, with its lower boundary $m_3$ fixed as 0.1. Dataset: MSR-VTT3k.}
    \label{fig:boundary}
\end{figure*}


\subsubsection{SNL or BNL?}

 As Table \ref{tab:BNL} shows, the best choice for SNL is $\ell_{tri}(x^+,q,q^-)$ (with the video as a pivot), scoring MIR of 0.408 on the original query set and 0.241 on the composed query set. The best choice of BNL is the joint use of $\ell_{bcl}(x^+,q,q^-)$ and $\ell_{bcl}(q,x^+,q^-)$, scoring MIR of 0.404 on the original and 0.274 on the composed. With both are better than the baseline w/o using any auxiliary loss, BNL is better than SNL in terms of the overall performance.
 
Interestingly, when comparing Row\#2 and Row\#3 in Table \ref{tab:BNL}, using the original query $q$ as the pivot for loss computation yields a model that is clearly more sensitive to the negation than that using the video $x^+$ as the pivot ($\Delta MIR$ 0.288 versus 0.021). A similar phenomenon can also be observed in BNL ($\Delta MIR$ 0.199 versus 0.030). Recall that the similarity between the query (video) pivot and the soft negative $q^-$ is computed in a text-to-text (video-to-text) manner. Hence, the results suggest that negation learning by text-to-text matching easily pushes the soft negative far away and adversely affects the learned common space. For this reason, we see that adding the upper-boundary constraint is important, increasing MIR on the composed set from 0.141 to 0.257 (Row\#3 versus Row\#5 in Table \ref{tab:BNL}).

\subsubsection{Influence of the Upper Boundary in  $\ell_{bcl}(q,x^+,q^-)$}

Following the above discussion about the upper boundary $m_4$, we further study its influence when using $\ell_{bcl}(q,x^+,q^-)$ as the auxiliary loss. As the performance curves in Fig. \ref{fig:boundary} show, while its impact on the original query set seems to be marginal, lowering its value obtains better performance on the composed query set. The result again justifies the necessity of BNL.

\section{Conclusions} \label{sec:concs}


To conquer the novel task of negation in text-to-video retrieval (nT2VR), we propose a new evaluation protocol together with a learning based method for negation-aware T2V model training. Our experiments on two re-purposed datasets, \ie MSR-VTT and VATEX, allow us to draw conclusions as follows. For the existing T2VR models evaluated in this paper, \ie W2VV++, SEA, CLIP and CLIP4Clip, they are all found to be unaware of negation in queries. Also, manipulating their retrieval results by boolean operations does not work. Negation learning by text-to-text matching easily pushes a soft negative description far away from its original description and adversely affects the learned common space. Bidirectional negation learning is thus necessary. Re-training the CLIP model by the proposed learning method  clearly improves its ability to handle queries with negation. In addition, its performance on the original benchmarks is also improved. We believe this work opens up new possibilities for multimedia retrieval.

\medskip

\textbf{Acknowledgments}. This work was supported by NSFC (No. 62172420, No. 62072463), BJNSF (No. 4202033), and Public Computing Cloud, Renmin University of China.



\bibliographystyle{ACM-Reference-Format}
\balance
\bibliography{mm2022-nt2vr}

\appendix

\section{Appendices}

\subsection{Rules for Query Construction}  \label{ssec:append-query}

\textbf{Negated Query Construction}.  Table \ref{tab:negated} shows example rules to generate negated queries.
\begin{table}[tbh!]
\caption{\textbf{Rule-based negated query construction}.}
\label{tab:negated}
\scalebox{0.65}{
\begin{tabular}{lll}
\toprule
\textbf{Original Query}& \textbf{Identified Word} (POS)  & \textbf{Negated Query Sample} \\ \hline
 \specialcell{Some guys are driving a car\\ and \underline{\textbf{met}} an accident in a road} & met (VBD) & \specialcell{Some guys are driving a car and \\ did not meet an accident in a road} \\ \hline
  \specialcell{ A cartoon alien character \underline{\textbf{finds}} \\ another character} & finds (VBZ) &   \specialcell{A cartoon alien character does not\\ find another character} \\ \hline
 \specialcell{ A man is \underline{\textbf{running}} around \\ and playing a guitar} & running (VBG) &  \specialcell{A man is not running around\\ and playing a guitar} \\  \hline
  \specialcell{A man \underline{\textbf{is}} running \\around and playing a guitar} & is (AUX) &  \specialcell{A man isn't running around\\ and playing a guitar} \\  \hline
  \specialcell{A father and son \underline{\textbf{are}} playing\\  with each others' hair} & are (AUX) &  \specialcell{A father and son aren't\\ playing with each others' hair} \\  \hline
  \specialcell{ A live concert \underline{\textbf{with}} a woman\\as the lead singer} & with  (ADP) &  \specialcell{A live concert without \\a woman as the lead singer} \\
\bottomrule
\end{tabular}
}
\end{table}


\begin{table*}[tbh!]
\caption{\textbf{Influence of the auxiliary-loss weight $\lambda_1$ on SNL. Each row corresponds to a specific CLIP model trained by weightedly adding
the corresponding auxiliary loss to the primary retrieval loss. Dataset: MSR-VTT3k}.}
\label{tab:abalation_snl}
\scalebox{0.7}{
\begin{tabular}{lrlrrrrlrrrrlrrrr}
\toprule
\multirow{2}{*}{\textbf{Auxiliary loss}} & \multicolumn{1}{r}{\multirow{2}{*}{\textbf{$\lambda_{1}$}}} &  & \multicolumn{4}{c}{\textbf{Original ($\uparrow$)}} & \multicolumn{1}{c}{} & \multicolumn{4}{c}{\textbf{Negated ($\uparrow$)}} & \multicolumn{1}{c}{} & \multicolumn{4}{c}{\textbf{Composed ($\uparrow$)}} \\ \cline{4-7} \cline{9-12} \cline{14-17} 
 & \multicolumn{1}{l}{} &  & \multicolumn{1}{l}{$R1$} & \multicolumn{1}{l}{$R5$} & \multicolumn{1}{l}{$R10$} & \multicolumn{1}{l}{$MIR$} &  & \multicolumn{1}{l}{$\Delta R1$} & \multicolumn{1}{l}{$\Delta R5$} & \multicolumn{1}{l}{$\Delta R10$} & \multicolumn{1}{l}{$\Delta MIR$} &  & \multicolumn{1}{l}{$R1$} & \multicolumn{1}{l}{$R5$} & \multicolumn{1}{l}{$R10$} & \multicolumn{1}{l}{$MIR$} \\ \hline
\textit{-- (Row\#4, Table 2)} & -- &  & 27.7 & 53.0 & 64.2 & 0.398 & \multicolumn{1}{r}{} & 0.5 & 1.1 & 1.1 & 0.008 & \multicolumn{1}{r}{} & 11.4 & 33.3 & 46.2 & 0.225 \\ \hline
\multirow{2}{*}{$\ell_{tri}(x^+, q, q^-)$} & 0.001 &  & \textbf{28.8} & \textbf{54.2} & \textbf{65.1} & \textbf{0.408} & \multicolumn{1}{r}{} & 1.8 & 2.5 & 2.5 & 0.021 & \multicolumn{1}{r}{} & \textbf{12.9} & \textbf{35.1} & \textbf{48.1} & \textbf{0.241} \\
 & 0.01 &  & 28.5 & 53.8 & 64.7 & 0.405 & \multicolumn{1}{r}{} & 11.3 & 16.7 & 17.5 & 0.135 & \multicolumn{1}{r}{} & 10.5 & 30.6 & 44.0 & 0.209 \\ \hline
\multirow{2}{*}{$\ell_{tri}(q, x^+, q^-)$} & 0.001 &  & 28.1 & 53.5 & 64.6 & 0.402 &  & 21.4 & 38.8 & 44.5 & 0.288 &  & 7.4 & 19.4 & 27.4 & 0.141 \\
 & 0.01 &  & 27.9 & 53.3 & 64.3 & 0.399 &  & \textbf{25.3} & \textbf{47.5} & \textbf{55.9} & \textbf{0.352} &  & 3.2 & 10.2 & 15.8 & 0.075 \\
 \bottomrule
\end{tabular}
}
\end{table*}
\begin{table*}[tbh!]
\caption{\textbf{Influence of the upper boundary ($m_2$ / $m_4$) and the auxiliary-loss weight $\lambda_2$ on BNL. Dataset: MSR-VTT3k}.}
\label{tab:abalation_bnl}
\scalebox{0.7}{
\begin{tabular}{lrrlrrrrlrrrrlrrrr}
\toprule
\multirow{2}{*}{\textbf{Auxiliary loss}} & \multicolumn{1}{r}{\multirow{2}{*}{\textbf{\begin{tabular}[c]{p{0.8cm}p{0.8cm}}Upper\\ boundary\end{tabular}}}} & \multicolumn{1}{r}{\multirow{2}{*}{\textbf{$\lambda_{2}$}}} &  & \multicolumn{4}{c}{\textbf{Original ($\uparrow$)}} & \multicolumn{1}{c}{} & \multicolumn{4}{c}{\textbf{Negated ($\uparrow$)}} & \multicolumn{1}{c}{} & \multicolumn{4}{c}{\textbf{Composed ($\uparrow$)}} \\ \cline{5-8} \cline{10-13} \cline{15-18} 
 & \multicolumn{1}{l}{} & \multicolumn{1}{l}{} &  & \multicolumn{1}{l}{$R1$} & \multicolumn{1}{l}{$R5$} & \multicolumn{1}{l}{$R10$} & \multicolumn{1}{l}{$MIR$} &  & \multicolumn{1}{l}{$\Delta R1$} & \multicolumn{1}{l}{$\Delta R5$} & \multicolumn{1}{l}{$\Delta R10$} & \multicolumn{1}{l}{$\Delta MIR$} &  & \multicolumn{1}{l}{$R1$} & \multicolumn{1}{l}{$R5$} & \multicolumn{1}{l}{$R10$} & \multicolumn{1}{l}{$MIR$} \\ \hline
\textit{-- (Row\#4, Table 2)} & -- & -- &  & 27.7 & 53.0 & 64.2 & 0.398 & \multicolumn{1}{r}{} & 0.5 & 1.1 & 1.1 & 0.008 & \multicolumn{1}{r}{} & 11.4 & 33.3 & 46.2 & 0.225 \\ \hline
\multirow{9}{*}{$\ell_{bcl}(x^+, q, q^-)$} & \multicolumn{1}{r}{$m_2$} & \multicolumn{1}{l}{} &  &  &  &  &  & \multicolumn{1}{r}{} &  &  &  &  & \multicolumn{1}{r}{} &  &  &  &  \\
 & 0.2 & 0.001 &  & 28.2 & 53.6 &  \textbf{64.6} & 0.403 &  & 0.9 & 1.3 & 1.3 & 0.011 &  & 12.1 & 35.2 & 47.9 & 0.236 \\
 & 0.2 & 0.01 &  & 27.8 & 53.0 & 64.0 & 0.397 &  & 3.3 & 4.6 & 4.6 & 0.038 &  & 11.3 & 30.9 & 43.1 & 0.214 \\ \cline{2-18} 
 & 0.3 & 0.001 &  & 27.8 & 53.1 & 64.3 & 0.398 & \multicolumn{1}{r}{} & 4.3 & 5.6 & 5.8 & 0.048 & \multicolumn{1}{r}{} & 13.9 & 37.3 & 50.2 & 0.256 \\  
 & 0.3 & 0.01 &  & 28.4 & 53.7 & \textbf{64.6} & 0.404 & \multicolumn{1}{r}{} & 6.5 & 9.0 & 9.2 & 0.074 & \multicolumn{1}{r}{} & 12.7 & 32.3 & 44.6 & 0.228 \\ \cline{2-18} 
 & 0.6 & 0.001 &  & 27.5 & 52.7 & 63.8 & 0.395 & \multicolumn{1}{r}{} & 2.7 & 3.7 & 3.6 & 0.030 & \multicolumn{1}{r}{} & 14.6 & \textbf{38.9} & \textbf{51.5} & \textbf{0.264} \\
 & 0.6 & 0.01 &  & \textbf{28.6} & \textbf{53.7} & 64.5 & \textbf{0.405} & \multicolumn{1}{r}{} & 9.5 & 13.9 & 14.4 & 0.112 & \multicolumn{1}{r}{} & 12.4 & 33.7 & 46.8 & 0.232 \\ \cline{2-18} 
 & 0.9 & 0.001 &  & 27.4 & 52.9 & 64.0 & 0.395 & \multicolumn{1}{r}{} & 3.5 & 5.3 & 5.1 & 0.041 & \multicolumn{1}{r}{} & 14.1 & 38.8 & 52.0 & 0.262 \\
 & 0.9 & 0.01 &  & 28.0 & 53.4 & \textbf{64.6} & 0.400 & \multicolumn{1}{r}{} & 11.7 & 18.0 & 19.3 & 0.143 & \multicolumn{1}{r}{} & 10.4 & 30.8 & 43.5 & 0.210 \\ \hline
\multirow{9}{*}{$\ell_{bcl}(q, x^+, q^-)$ } & \multicolumn{1}{r}{$m_4$} & \multicolumn{1}{l}{} &  & \multicolumn{1}{l}{} & \multicolumn{1}{l}{} & \multicolumn{1}{l}{} & \multicolumn{1}{l}{} &  & \multicolumn{1}{l}{} & \multicolumn{1}{l}{} & \multicolumn{1}{l}{} & \multicolumn{1}{l}{} &  & \multicolumn{1}{l}{\textbf{}} & \multicolumn{1}{l}{} & \multicolumn{1}{l}{} & \multicolumn{1}{l}{} \\
 & 0.2 & 0.001 &  & 28.0 & 53.3 & 64.2 & 0.400 &  & 15.8 & 26.0 & 28.1 & 0.199 &  & \textbf{14.8} & 36.1 & 50.4 & 0.257 \\
 & 0.2 & 0.01 &  & 28.2 & 53.5 & \textbf{64.6} & 0.402 &  & 11.2 & 16.9 & 17.3 & 0.133 &  & 10.0 & 29.4 & 40.6 & 0.198 \\ \cline{2-18} 
 & 0.3 & 0.001 &  & 28.3 & 53.5 & 64.3 & 0.402 &  & 17.0 & 27.5 & 30.0 & 0.213 &  & 11.8 & 30.6 & 42.8 & 0.216 \\
 & 0.3 & 0.01 &  & 28.3 & 53.4 & 64.2 & 0.402 &  & 15.5 & 24.5 & 26.1 & 0.191 &  & 7.4 & 23.7 & 33.6 & 0.159 \\ \cline{2-18} 
 & 0.6 & 0.001 &  & 28.3 & 53.5 & 64.3 & 0.402 &  & 23.1 & 42.6 & 49.7 & 0.316 &  & 5.6 & 14.3 & 19.8 & 0.108 \\
 & 0.6 & 0.01 &  & 28.4 & 53.4 & 64.2 & 0.402 &  & \textbf{26.6} & \textbf{49.3} & 58.1 & \textbf{0.368} &  & 1.0 & 4.6 & 8.8 & 0.039 \\ \cline{2-18} 
 & 0.9 & 0.001 &  & 27.2 & 52.9 & 64.0 & 0.394 &  & 21.3 & 38.7 & 44.4 & 0.288 &  & 7.1 & 19.5 & 28.3 & 0.139 \\
 & 0.9 & 0.01 &  & 28.4 & 53.6 & \textbf{64.6} & 0.404 &  & 26.4 & 49.1 & \textbf{58.2} & 0.367 &  & 1.1 & 4.5 & 8.0 & 0.039 \\
 \bottomrule
\end{tabular}
}
\end{table*}
\textbf{Composed Query Generation}. 
We use templates to generate composed queries. According to whether the pronoun of subject can be determined, we use two sets of templates. For example, given < kids, do A, do B>, we first get the pronoun of `kids', and then randomly choose one of following queries:\\
{\footnotesize
$\bullet$ Kids do A and they don't do B. \\
$\bullet$  Kids don't do B and they do A. \\
$\bullet$  Kids doing A and not doing B. \\
$\bullet$  Kids not doing B and they doing A. \\
$\bullet$  Kids are doing A and not doing B. \\
$\bullet$  Kids are not doing B and they are doing A.} 

When the subject is in third-person singular and gender cannot be determined, the pronoun of the subject is unknown. We then use a set of slightly different templates. For example, given <A kid, do A, do B>, we randomly choose one of the following queries: \\
{\footnotesize
$\bullet$  A kid does A and doesn't do B. \\
$\bullet$  A kid doesn't do B while does A. \\
$\bullet$  A kid doing A and not doing B. \\
$\bullet$  A kid not doing B while doing A. \\
$\bullet$  A kid is doing A and not doing B. \\
$\bullet$  A kid is not doing B while doing A. }

\subsection{Hyper-parameter Evaluation}
\label{ssec:append-hyper}

\textbf{Influence of the auxiliary-loss weight $\lambda_1$ on SNL}.
As Table \ref{tab:abalation_snl} shows, setting the auxiliary-loss weight $\lambda_1$ as 0.001 yields better performance on both original and composed query set. Setting $\lambda_1$ as 0.01 make model more sensitive to negation but at the cost of undesired performance drop on the two other query sets.

\textbf{Influence of the upper boundary ($m_2$ / $m_4$) and the auxiliary-loss weight $\lambda_2$ on BNL}. 
As Table \ref{tab:abalation_bnl} shows, upper boundary and auxiliary learning weight jointly influence negation learning task. Generally, with looser upper boundary and larger auxiliary-loss weight, model exhibits more sensitiveness to negation, but their performance not necessarily increases on the composed and original query sets. Their influence over original query set is smaller than composed query set. Using $\ell_{bcl}(x^+, q, q^-)$ alone (with the video as a pivot), top performance on composed query set is achieved by setting $m_2$ as 0.6 and $\lambda_2$ as 0.001. Using $\ell_{bcl}(q, x^+, q^-)$ alone (with the original query as a pivot), top performance on composed query set is achieved by setting $m_4$ as 0.2 and $\lambda_2$ as 0.001. The result suggests that setting auxiliary-loss weight as 0.001 is more appropriate, meanwhile using the query pivot requires a tighter upper boundary than using the video pivot.

\subsection{NL for Text-to-Image Retrieval (T2IR)}
To see to what extent can our findings be generalized to the image domain, we reproduce our research on Flickr30k \cite{young-etal-2014-image} and MS-COCO \cite{Lin2014MicrosoftCC}, with results shown in Table \ref{tab:flickr} and \ref{tab:coco}.  We simply adopt the same hyper-parameters as we have used for T2VR, which could be suboptimal for T2IR. Again, we observe that CLIP-\emph{bnl} outperforms the baselines on the composed query set with a clear margin, showing the viability of the proposed negation learning method. Although our paper is targeted at T2VR, the proposed negative learning method also works for T2IR with negation.

\begin{table}[tbh!]
\caption{Results on re-purposed Flickr30k (data split: \cite{young-etal-2014-image}).}
\label{tab:flickr}
\scalebox{0.6}{
\begin{tabular}{lrrrrlrrrrlrrrr}
\toprule
\multirow{2}{*}{\textbf{Models}} & \multicolumn{4}{c}{\textbf{Original ($\uparrow$)}} & \multicolumn{1}{c}{\textbf{}} & \multicolumn{4}{c}{\textbf{Negated ($\uparrow$)}} & \multicolumn{1}{c}{\textbf{}} & \multicolumn{4}{c}{\textbf{Composed ($\uparrow$)}} \\ \cline{2-5} \cline{7-10} \cline{12-15} 
 & $R1$ & $R5$ & $R10$ & $MIR$ & \multicolumn{1}{r}{} & $\Delta R1$ & $\Delta R5$ & $\Delta R10$ & $\Delta MIR$ & \multicolumn{1}{r}{} & $R1$ & $R5$ & $R10$ & $MIR$ \\ \hline
CLIP & 59.0 & 84.6 & 91.0 & 0.705 &  & 2.7 & 2.0 & 1.2 & 0.024 &  & 18.9 & 41.2 & 56.0 & 0.303 \\
CLIP* & \textbf{75.1} & \textbf{93.3} & \textbf{96.1} & \textbf{0.832} &  & 1.2 & 0.4 & 0.2 & 0.009 &  & 23.4 & 50.9 & 65.4 & 0.365 \\
CLIP (boolean) & -- & -- & -- & --  &  & 52.3 & 69.4 & 71.5 & 0.594 &  & 8.5 & 20.1 & 28.3 & 0.150 \\
CLIP*(boolean) & -- & -- & -- & --  & \textbf{} & \textbf{65.2} & \textbf{72.2} & \textbf{70.2} & \textbf{0.677} &  & 16.1 & 41.2 & 51.6 & 0.278 \\
CLIP-\emph{bnl} & 74.8 & 93.1 & 96.2 & 0.829 &  & 18.5 & 10.7 & 6.8 & 0.149 & \textbf{} & \textbf{26.4} & \textbf{55.7} & \textbf{70.8} & \textbf{0.398} \\ 
 \bottomrule
 \end{tabular}
}
\end{table}

\begin{table}[tbh!]
\caption{Results on re-purposed MS-COCO (data split: \cite{Karpathycoco}). }
\label{tab:coco}
\scalebox{0.6}{
\begin{tabular}{lrrrrlrrrrlrrrr}
\toprule
\multirow{2}{*}{\textbf{Models}} & \multicolumn{4}{c}{\textbf{Original}} & \multicolumn{1}{r}{\textbf{}} & \multicolumn{4}{c}{\textbf{Negated}} & \multicolumn{1}{r}{\textbf{}} & \multicolumn{4}{c}{\textbf{Composed}} \\ \cline{2-5} \cline{7-10} \cline{12-15} 
 & $R1$ & $R5$ & $R10$ & $MIR$ & \multicolumn{1}{r}{} & $\Delta R1$ & $\Delta R5$ & $\Delta R10$ & $\Delta MIR$ & \multicolumn{1}{r}{} & $R1$ & $R5$ & $R10$ & $MIR$ \\ \hline
CLIP & 28.8 & 54.1 & 65.0 & 0.408 &  & 1.6 & 2.6 & 2.3 & 0.020 &  & 11.8 & 28.9 & 40.6 & 0.210 \\
CLIP* & \textbf{45.6} & \textbf{72.8} & \textbf{82.3} & \textbf{0.579} &  & 3.2 & 1.9 & 1.7 & 0.025 &  & 14.0 & 38.2 & 52.7 & 0.261 \\
CLIP (boolean) & -- & -- & -- & -- &  & 24.3 & 43.4 & 50.1 & 0.330 &  & 7.0 & 17.7 & 24.9 & 0.127 \\
CLIP*(boolean) & -- & -- & -- & -- & \textbf{} & \textbf{39.5} & \textbf{57.1} & \textbf{61.6} & \textbf{0.467} &  & 12.8 & 32.0 & 42.5 & 0.222 \\
CLIP-\emph{bnl} & 44.7 & 71.8 & 81.4 & 0.570 &  & 18.0 & 18.7 & 16.1 & 0.176 & \textbf{} & \textbf{17.3} & \textbf{43.0} & \textbf{56.1} & \textbf{0.298} \\ 
 \bottomrule
 \end{tabular}
}
\end{table}

\end{document}